\documentstyle[12pt]{article}
\begin{document}
\textwidth=159mm
\textheight=220mm
\setlength{\leftmargin}{-1.0in}

\def\sst#1{{\scriptscriptstyle #1}}
\def\fft#1#2{{#1 \over #2}}
\def\ft#1#2{{\scriptstyle {#1 \over #2}}}
\def\diag{{\,\rm diag}}
\def\dfi{{\del\varphi}}
\def\teff{{T^{\rm eff}}}
\def\cmin{{c_{\rm min}}}
\def\oneone{\rlap 1\mkern4mu{\rm l}}
\def\e{\epsilon}
\def\ve{\varepsilon}
\def\del{\partial}
\def\p{\psi}
\def\g{\gamma}
\def\a{\alpha}
\def\b{\beta}
\def\d{\delta}
\def\s{\sigma}
\def\o{\omega}
\def\ra{\rightarrow}
\def\h{\hat}
\def\vp{\varphi}


\topmargin 0pt
\oddsidemargin 5mm
\begin{titlepage}
\begin{flushright}
CTP TAMU-15/97\\
hep-th/9703063\\
\end{flushright}
\vspace{1.5truecm}
\begin{center}
{\bf {\Large $D=9$ supergravity and $p$-brane solitons}}
\vspace{1.5truecm}

{\large N. Khviengia and Z. Khviengia}
\vspace{1.1truecm}

{\small Center for Theoretical Physics, Texas A\&M University,
                College Station, TX 77843-4242} \vspace{1.1truecm}


\end{center}
\vspace{1.0truecm}

\begin{abstract}
\vspace{1.0truecm}
We construct the $N=2$, $D=9$ supergravity theory up to the 
quartic fermionic terms and derive the supersymmetry 
transformation rules for the fields modulo cubic fermions. 
We consider a class of $p$-brane solutions of this theory, the 
stainless $p$-branes, which cannot be isotropically oxidized into 
higher dimensions. The new stainless elementary membrane 
and elementary particle solutions are found.  It is explicitly 
verified that these solutions preserve half of the supersymmetry.

\end{abstract}
\end{titlepage}
\newpage
\pagestyle{plain}
\section{Introduction}

The recent progress in the understanding of M-theory revived the interest 
in the supergravity theories  in diverse dimensions.  Much work has been 
done in constructing and classifying the $p$-brane solutions  to these 
theories. However, the extended supergravity in nine dimensions has not 
yet been fully constructed   and  investigated. Purely bosonic $N=2$, 
$D=9$ action  in the context of type-II S- and T-duality  symmetries has 
been discussed in \cite{berg2,das}, but the full action of the theory and  
the supersymmetry transformation rules has not  appeared in the 
literature. The goal of this paper is  to fill this gap and present an 
explicit construction of the $N=2$, $D=9$ supergravity  up to the quartic 
fermionic terms and provide an exhaustive classification of stainless 
$p$-brane solutions of this theory.  

With few exceptions, the lower dimensional supergravity theories can be 
obtained by dimensional reduction of the eleven dimensional 
Cremmer-Julia-Scherk (CJS) supergravity \cite{CJS}.  Some examples of  such 
exceptions in $D=10$  are provided by type IIB and massive type IIA 
supergravity theories \cite{rom,berg1}. Dimensional reduction of CJS 
action, apart from massless supergravities, can also give massive 
theories in $D\leq 8$ \cite{SS,LPmass}. As one descends through the 
dimensions to obtain lower dimensional supergravities, a plethora of 
isotropic $p$-brane solutions arises \cite{duff,NZ,LPmax}.  Solutions 
of the dimensionally reduced theory are also solution of the 
higher-dimensional theory, however, in higher dimension these solutions 
may or may not exhibit isotropicity.  The solutions which cannot be 
isotropically lifted to the higher dimension and, therefore, cannot be 
viewed as $p$-brane solutions in the higher dimension, are called 
stainless solutions \cite{LPSS}.  The extended $N=2$ supergravity in 
nine dimensions can be truncated to $N=1$ supergravity  whose 
stainless solutions consist of an elementary particle and a solitonic 
5-brane  \cite{LPSS}.  It turns out that, whilst the elementary particle 
remains stainless in $N=2$ theory, the solitonic 5-brane becomes 
rusty and can be obtained from the type IIA $D=10$ solitonic 6-brane. 
We explain this phenomenon by  employing a two-scalar 5-brane 
solution of $N=2$ supergravity.

The paper is organised as follows. In sections 2, using the ordinary 
Scherk-Schwarz dimensional reduction procedure \cite{SS}, we 
obtain the bosonic Lagrangian of $N=2$, $D=9$ supergravity 
theory by dimensionally reducing the eleven-dimensional 
CJS Lagrangian from eleven directly to nine dimensions. It is of 
interest to note that in $D=9$, unlike $D=8$ case \cite{sez}, 
nontrivial group manifolds do not arise. In section 3, we perform an 
analogous dimensional reduction for fermions and obtain
the fermionic part of the nine-dimensional supergravity up to 
quartic fermions. The supersymmetry transformation rules for 
the bosonic and fermionic fields, modulo trilinear fermions, are 
derived in secion 4. In section 5, stainless solutions to the obtained 
$N=2$, $D=9$ supergravity are analysed. New stainless 
elementary particle and elementary membrane solutions are found, 
and it is shown that they preserve half of the supersymmetry. A 
stainless solitonic 6-brane is also discussed. It is noted that if one 
is to include type IIB chiral supergravity into consideration,
the solitonic 6-brane and the elementary membrane solutions 
become rusty and can be treated as descendants of the type IIB 
solitonic 7-brane and self-dual 3-brane in $D=10$.

\section{Bosonic Sector}

The bosonic part of the $N=1,\  D=11$ supergravity Lagrangian is 
given by \cite{CJS}

\begin{equation}
{\cal L}=\fft{\h e}{4\kappa^2}\h R(\omega)-\fft{\h e}{48}
\h F_{\h \mu\h \nu\h \rho\h \s}
\h F^{\h \mu\h \nu\h \rho\h \s}
+\fft{2\kappa}{(12)^4}\ve^{\h \mu_1\cdots\h \mu_{11}}
\h F_{\h \mu_1\cdots}\h F_{\cdots}
\h A_{\cdots\h \mu_{11}}\ ,\label{CJS}
\end{equation}
where $\h e=det ({\h e}_{\h \mu}^{~\h r})$, and   
$\h F_{\h \mu\h \nu\h \rho\h \s}$ is the field 
strength associated with the gauge field 
${\h A}_{\h \nu\h \rho\h \s}$
\begin{equation}
\h F_{\h \mu\h \nu\h \rho\h \s}=4\del_{[\h \mu}
{\h A}_{\h \nu\h \rho\h \s]}\ .
\end{equation}
Here $\h \mu,\h \nu,\cdots=0,1,\cdots10$ are the world indices 
and the square brackets represent the antisymmetrization with the 
unit strength.  We take the metric to be mostly positive 
and perform  the dimensional reduction in the  space-like direction.

The Riemann tensor and the curvature scalar are as follows
\begin{eqnarray}
\h R^{\h r}_{~\h s\h \mu\h \nu}&=&\del_{\h \mu}
{~\h \o}_{\h \nu~\h s}^{~\h r}
+{\h \o}_{\h \mu~\h t}^{~\h r}\ {\h \o}_{\h \nu~\h s}^{~\h t}
-(\h \mu\leftrightarrow {\h \nu}), \\
{\h R} &=&\h e_{\h r}^{~\h \mu} {\h e}_{\h \nu}^{~\h t} 
  {\h R}^{\h r}_{~\h s\h \mu\h \nu}\eta^{\h s\h t}\ ,
\end{eqnarray}
where $\h r,\h s,\cdots=0,1,\cdots,10$ denote the eleven-
dimensional flat indices. 

The spin connection is defined by
\begin{equation}
\h \omega_{\h r\h s}^{~~\h t}=\ft{1}{2} (\h e_{\h r}^{~\h \mu}
\h e_{\h s}^{~\h \nu}-
\h e_{\h s}^{~\h \mu}\h e_{\h r}^{~\h \nu})\partial_{\h \mu}
\h e_{\h \nu}^{~\h t}\ .
\end{equation}

We shall now perform the ordinary Scherk-Schwarz dimensional 
reduction of CJS Lagrangian (\ref{CJS}) directly to $D=9$ 
dimensions.  We begin by dimensionally reducing the Einstein-
Hilbert term  in (\ref{CJS}). It is convenient first to consider 
reduction from an arbitrary $D+d$ to $D$ dimensions, 
and then apply obtained formulas to our case where $D+d=11$ 
and $D=9$. 

The Lorentz invariance of the supergravity theory in $D+d$ 
dimensions, enables one to cast the vielbein into the triangular 
form 
\begin{equation}
{\h e}_{\h \mu}^{~\h r}=
\pmatrix{{\h e}_\mu^{~r} &{\h {\cal A}}_\mu^i\cr 
0 & {\h e}_\alpha^{~i}\cr}\ . \label{veil1}
\end{equation}
Here  $\mu,r=0,1,\cdots,D-1$;  $\alpha,i=1,\cdots,d$, 
the hatted indices belong to $(D+d)$-dimensional space, and 
${\h {\cal A}}_\mu^i$ are the vector gauge fields that give rise 
to 2-index field strengths in the dimensionally reduced theory. 

Then the inverse veilbein is given by
\begin{equation}
{\h e}^{~\h \mu}_{\h r}=
\pmatrix{{\h e}^{~\mu}_r & -{\h {\cal A}}_r^\alpha \cr 
0 & {\h e}^{~\alpha}_i\cr}\ , \label{veil2}
\end{equation}
where internal indices are raised and lowered by the metric
\begin{equation} 
{\h g}_{\alpha\beta}={\h e}^{~i}_\alpha {\h e}^{~j}_\beta
\delta_{ij}\ .
\end{equation}
The components of  the spin-connection 
${\h \omega}_{\h r\h s\h t}$  are given by \cite{SS}
\begin{eqnarray}
\h \omega_{rsj}&=&\ft{1}{2}\h e_{j\alpha}
                {\h {\cal F}}^\alpha_{rs}\ , \nonumber\\
\h \omega_{rij}&=&\ft{1}{2}\h e^{~\alpha}_i
           \h e^{~\mu}_r\partial_\mu\h e_{j\alpha}-
                              (i\leftrightarrow j)\ , \nonumber\\
\h \omega_{jrs}&=&-\h \omega_{rsj}\ , \label{om1}\\
\h \omega_{irj}&=&-\ft{1}{2}\h e^{~\alpha}_i 
           \h e^{~\beta}_j \h e^{~\mu}_r
           \partial_\mu \h g_{\alpha\beta}\ , \nonumber
\end{eqnarray}
where $\omega_{rst}$ is the torsion-free spin connection in 
$D$ dimensions and 
${\h {\cal F}}_{rs}^\alpha=e_r^{~\mu} e_s^{~\nu} 
{\h {\cal F}}_{\mu\nu}^\alpha$ with
\begin{equation}
{\h {\cal F}}_{\mu\nu}^\alpha=\partial_\mu
        {\h {\cal A}}_\nu^\alpha-
       \partial_\nu{\h {\cal A}}_\mu^\alpha.
\end{equation}
Using  (\ref{om1}),  the following general formula for 
the dimensional reduction of the Einstein-Hilbert Lagrangian can 
be obtained \cite{das,SS}:
\begin{eqnarray}
\int d^{D+d} x {\h e}{\h R} &=& \int d^Dx(det
           {\h e}^{~r}_\mu)\delta
\Big[R-\ft14 {\h g}_{\alpha \beta}{\h {\cal F}}^{\mu\nu\alpha}
         {{\h {\cal F}}_{\mu\nu}^\beta}\\
&&+ \ft14{\h g}^{\mu\nu}{\partial_\mu} {\h g}_{\alpha\beta}
          {\partial_\nu} {\h g}^{\alpha\beta}
+{\h g}^{\mu\nu}{\partial_\mu ln\delta}\ {\partial_\nu ln\delta}
    \Big]\ , \label{EH1}
\end{eqnarray}
where $\delta=det({\h e}^{~i}_\alpha)$ and $R$ is a Ricci scalar 
in $D$ dimensions.

We perform the following rescaling \cite{SS}
\begin{eqnarray}
\h e^{~r}_\mu&=&\delta^\gamma e^{~r}_\mu\ , \nonumber\\
\h e^{~i}_\alpha&=&\delta^{\frac{1}{d}}L_\alpha^{~i}\ ,  
     \label{resc}\\
{\h {\cal A}}_\mu^\alpha&=&2\kappa{\cal A}_\mu^\alpha\ , 
     \nonumber
\end{eqnarray}
where $L_\alpha^{~i}$ is a unimodular matrix 
$detL_\alpha^{~i}=1$, and $\gamma$ is
a free parameter that determines an exponential prefactor of the 
Einstein-Hilbert term in a lower dimensional theory.

 As a result of rescaling, the vielbeins are brought to the following
form
\begin{equation}
{\h e}_{\h \mu}^{~\h r}=
\pmatrix{\delta^\gamma e_\mu^{~r} &2\kappa
       \delta^{\frac{1}{d}}{\cal A}_\mu^\alpha L_\alpha^{~i}\cr 
0 &\delta^{\frac{1}{d}} L _\alpha^{~i}\cr},\qquad
{\h e}^{~\h \mu}_{\h r}=
\pmatrix{\delta^{-\gamma}e^{~\mu}_r & -2\kappa
         \delta^{-\gamma}{\cal A}_r^\alpha \cr 
0 &\delta^{-\frac{1}{d}} L^{~\alpha}_i\cr} . \label{veil}
\end{equation}

Under the rescaling of the type (\ref{resc}), the metric in D 
dimensions rescales as 
$\h g_{\mu\nu}=\delta^{2\gamma}g_{\mu\nu}$, and a Ricci 
scalar changes as
\begin{equation}
R\ra\delta^{-2\gamma}\Big[R-2\gamma (D-1)g^{\mu\nu}
         {\nabla_\mu}
        {\nabla_\nu}ln\delta-\gamma^2 (D-1)(D-2)g^{\mu\nu}
        {\nabla_\mu}
        ln\delta\ {\nabla_\nu}ln\delta\Big]. \label{hR}
\end{equation}
Using (\ref{EH1}) and (\ref{hR}), we obtain the expression for 
the dimensional reduction of the Einstein-Hilbert action  from 
$D+d$ to $D$ dimensions (for $d>1$) that generalises the
result of ref. \cite{SS} for the arbitrary value of parameter 
$\gamma$
\begin{eqnarray}
\int d^{D+d} x {\h e}{\h R} &=& \int d^Dxe
       \delta^{\gamma(D-2)+1}
\Big[R- \kappa^2\delta^{-2\gamma +\frac{2}{d}}
       g_{\alpha \beta}{\cal F}^{\mu\nu\alpha}
{{\cal F}_{\mu\nu}^\beta}\cr
&&+
\beta (\gamma,D,d) g^{\mu\nu}{\partial_\mu}ln\delta 
          {\partial_\nu}ln\delta
     +\ft14g^{\mu\nu}{\partial_\mu} {g_{\alpha\beta}}
		{\partial_\nu}{g^{\alpha\beta}}\Big], \label{EH2}
\end{eqnarray}
where 
\begin{equation}
\b\equiv\gamma^2(D-1)(D-2)+2\gamma(D-1)+1-\frac{1}{d}
\end{equation}
and $g_{\alpha\beta}=L_\alpha^{~i}L_\beta^{~j}\delta_{ij}$, 
$e=det(e_\mu^{~r})$ and 
$\h e=\delta^{\gamma D+1}e$.  One, of course, is free to choose not to rescale 
the veilbein as (\ref{resc}) and work with the unrescaled veilbein 
given by (\ref{veil1}). In that case, the expressions above would have to 
be taken with $\gamma\ra 0$ and $d\ra\infty$.

Using (\ref{resc}), the components of the spin connection are 
obtained \cite{SS}
\begin{eqnarray}
\h \omega_{rst}&=&\delta^{-\gamma}\Big[\omega_{rst}
     +  \gamma\eta_{rs}\partial_t ln\delta-\g\eta_{rt} 
      \partial_s ln\delta\Big]\ , \nonumber\\
\h \omega_{rsj}&=&\kappa\delta^{-2\gamma+\frac{1}{d}}
      {\cal F}_{rsj}\ , \nonumber\\
\h \omega_{rij}&=&\delta^{-\gamma}Q_{rij}\ , \nonumber\\
\h \omega_{jrs}&=&-\h \omega_{rsj}\ , \\ 
\h \omega_{ijr}&=&\delta^{-\gamma}\left(2P_{rij}
                             +\ft1d\delta_{ij}\partial_r ln\delta\right)\ , 
          \nonumber\\
\h \omega_{ijk}&=&0\ , \label{omega} \nonumber
\end{eqnarray}
where  $P_{rij}$ is symmetric and traceless and 
$Q_{rij}$ is antisymmetric and defined as
\begin{equation}
P_{r ij}=\ft12 L_i^\alpha \partial_r L_{\alpha j}
       +(i\leftrightarrow j),\quad
Q_{\mu ij}=\ft12 L_i^\alpha \partial_\mu L_{\alpha j}
    -(i\leftrightarrow j)\ .
\end{equation}

We now turn to CJS Lagrangian (\ref{CJS}) and dimensionally 
reduce it to nine dimensions by applying the formulas 
(\ref{EH2})-(\ref{omega}) with $D+d=11$ and $D=9$.  One 
can identify nine-dimensional gauge fields  in terms of the eleven-
dimensional gauge field as follows
\begin{eqnarray}
A_{\mu\nu\rho}&=&{\h A}_{\mu\nu\rho}\ , \nonumber\\
A_{\mu\nu\alpha}&=&{\h A}_{\mu\nu\alpha}\ , \nonumber\\
A_{\mu\alpha\beta}&=&{\h A}_{\mu\alpha\beta}\  . \label{d9A}
\end{eqnarray}

It should be remarked here that there is an element of 
arbitrariness in defining the gauge field in $D=9$. Let us recall 
that CJS Lagrangian is invariant under the general coordinate 
transformation \cite{SS}. Upon compactification, this symmetry 
becomes $D=9$ general coordinate transformation and a set of the 
$[U(1)]^2$ reparametrization transformations.
Denoting a parameter in $D=11$ by $\xi^{\h \mu}$, one
can show that under the $\xi^\alpha$-reparametrization 
transformation, the gauge fields defined in (\ref{d9A}) transform 
noncovariantly   involving derivatives of the parameter 
$\xi^\alpha$. Nevertheless, the supersymmetry transformation 
rules and the dimensional reduction procedure is somewhat 
simplified by this choice \cite{duff,LPSS}.  In order to obtain  
results in terms of the covariant gauge fields, used in \cite{SS},  one 
has to change conventions by identifying the nine-dimensional 
gauge fields as $B_{rst}=\delta^{3\gamma}{\h A}_{rst}$, 
$B_{rsi}=\delta^{2\gamma}{\h A}_{rsi}$ and
$B_{rij}=\delta^{\gamma}{\h A}_{rij}$. This amounts to 
the following redefinitions of the fields: 
\begin{eqnarray}
B_{\mu\nu\rho}&=&A_{\mu\nu\rho}-6\kappa 
           {\cal A}^i_{[\mu}A_{\nu\rho]i}
      +12\kappa^2{\cal A}^i_{[\mu}{\cal A}^j_\nu A_{\rho]ij}\ , 
         \nonumber\\
B_{\mu\nu i}&=&A_{\mu\nu i}-4\kappa {\cal A}^j_{[\mu}
         A_{\nu i]j}\ , \label{BA}\\
B_{\mu ij}&=&A_{\mu ij}\ . \nonumber
\end{eqnarray}

Dimensionally reducing the field strength, we find
\begin{eqnarray}
{\h F}_{rstu}&=&\delta^{-4\gamma}\left(F_{rstu}-2kF_{rsti}
{\cal A}^i_u+4k^2F_{rsij}{\cal A}_t^i{\cal A}_u^j\right)
\equiv\delta^{-4\gamma} F'_{rstu}\ , \nonumber\\
{\h F}_{rsti}&=&\delta^{-3\gamma-\ft12}
       \left(F_{rsti}-4kF_{rsji}
{\cal A}^j_t\right)\equiv\delta^{-3\gamma-\ft12}F'_{rsti}\ , 
      \nonumber\\
{\h F}_{rsij}&=&\delta^{-2\gamma-1}F_{rsij}\ , \label{hF}\\
{\h F}_{rijk}&=&0\ , \nonumber\\
{\h F}_{ijkl}&=&0\ . \nonumber
\end{eqnarray}

Above rules reflect the fact that the field strengths in $D=9$ do 
not transform covariantly under the U(1) transformations 
arising from eleven-dimensional general coordinate 
transformation. This $U(1)$ reparametrization invariance 
should not be confused with another $U(1)$ symmetry which 
is a gauge symmetry of the antisymmetric field  in $D=9$. 

In $D=11$, the $U(1)$ gauge transformation  is given by 
\cite{SS}
\begin{equation}
\delta_{\sst {\Lambda}}{\h A}_{{\h \mu}{\h \nu}{\h \rho}}=
	3\del_{[{\h \mu}}\Lambda_{{\h \nu}{\h \rho}]}\ ,
\end{equation}
which upon reduction  gives the following $U(1)$ gauge 
transformations \\ 
in $D=9$: 
\begin{eqnarray}
\delta_{\sst {\Lambda}} A_{\mu\nu\rho}&=&
       3\del_{[\mu}\Lambda_{\nu\rho ]}\ , \nonumber\\
\delta_{\sst {\Lambda}} A_{\mu\nu\alpha}&=&
      2\del_{[\mu}\Lambda_{\nu ]\alpha}\ , \label{gauge}\\
\delta_{\sst {\Lambda}} A_{\mu\alpha\beta}&=&
     \del_{\mu}\Lambda_{\alpha\beta}\ . \nonumber
\end{eqnarray}

Notice that since we are performing the ordinary dimensional 
reduction, none of the fields and parameters in $D=9$ depend 
on extra compactification coordinates, in other words, all $\del_\alpha$ 
derivatives are identically zero. 

Turning to the kinetic term for 
the antisymmetric tensor field in (\ref{CJS}) and performing the 
straightforward reduction of the field strengths using (\ref{hF}), 
we obtain 
\begin{equation}
-\ft{e}{48} \delta^{9\gamma+1}\Big(\delta^{-8\gamma} 
    F'_{\mu\nu\rho\sigma}F'^{\mu\nu\rho\sigma}
+4\delta^{-6\gamma-1}g_{\alpha\beta}
   {F'_{\mu\nu\rho}}^{~\alpha} F'^{\mu\nu\rho\beta}
+6\delta^{-4\gamma-2} F'_{\mu\nu\alpha\beta}
   F'^{\mu\nu\alpha\beta}\Big). \label{F2}
\end{equation}

The interaction term in the bosonic Lagrangian can also be 
dimensionally reduced by decomposing summation in the hatted 
indices 
\begin{eqnarray}
&&\fft{2\kappa}{(12)^4}\ve^{\mu_1\cdots\mu_9}
     \ve^{\alpha\beta}\Big(
3F_{\mu_1\mu_2\mu_3\mu4}F_{\mu_5\mu_6\mu_7\mu_8}
      A_{\mu_9 \alpha\beta}
+24F_{\mu_1\mu_2\mu_3\mu_4}
       F_{\mu_5\mu_6\mu_7 \alpha}A_{\mu_8\mu_9\beta}\cr
&& \ \ \ \ \ \  - 16F_{\mu_1\mu_2\mu_3\alpha}
      F_{\mu_4\mu_5\mu_6\beta}A_{\mu_7\mu_8\mu_9}
+12F_{\mu_1\mu_2\mu_3\mu_4}F_{\mu_5\mu_6\alpha\beta} 
     A_{\mu_7\mu_8\mu_9}\Big), 
      \label{FFA}
\end{eqnarray}
where the totally antisymmetric tensor in nine dimensions is 
defined as $\ve^{\mu_1\cdots\mu_9}\ve^{\alpha\beta}=
\ve^{\mu_1\cdots\mu_9\alpha\beta}$.

In order to have the canonical normalization of  a scalar field 
kinetic term in (\ref{EH2}), we introduce a dilaton in D dimensions
\begin{equation}
\d=e^{\sqrt{-\ft2{\b}}\kappa\phi}\label{dlt}
\end{equation}
and choose  $\gamma =-\frac{1}{D-2}$ to bring  the Einstein-Hilbert 
term of the reduced theory to the standard form. A 
generalized method of dimensional reduction for this particular 
choice of parameters was discussed in \cite{SS}.

Combining (\ref{EH2}), (\ref{F2}), (\ref{FFA}), and choosing 
$\gamma=-\ft{1}{D-2}=-\ft{1}{7}$ and $\delta=
e^{\frac{2\!\sqrt{7}}{3}\kappa\phi}$,
 we get the bosonic action of the $N=2$, $D=9$ supergravity 
\cite{berg2,das}
\begin{eqnarray}
S &=& \int d^9xe
          \Big[\ft{1}{4\kappa^2}R- \ft{1}{4}e^{\ft{6}{\!\sqrt{7}}
        \kappa\phi}g_{\alpha \beta}
          {\cal F}^{\mu\nu\alpha}{{\cal F}_{\mu\nu}^\beta}
		-\ft{1}{2}({\partial_\mu}\phi)^2-\ft{1}{4\kappa^2}
          P_{\mu ij}P^{\mu ij}\cr
&&-\ft{1}{48}e^{\frac{4}{\!\sqrt{7}}\kappa\phi}
        F'_{\mu\nu\rho\sigma}F'^{\mu\nu\rho\sigma}
    -\ft{1}{12}e^{-\frac{2}{\!\sqrt{7}}\kappa\phi}
       g_{\alpha\beta}{F'_{\mu\nu\rho}}^{~\alpha} 
                 F'^{\mu\nu\rho\beta}\cr
&& -\ft{1}{8}e^{-\frac{8}{\!\sqrt{7}}\kappa\phi} 
        F_{\mu\nu\alpha\beta}F^{\mu\nu\alpha\beta}
              +{\cal L}_{\sst{FFA}}\Big]\ , \label{Sbos1}
\end{eqnarray}
where ${\cal L}_{\sst{FFA}}$ is given by (\ref{FFA}).

The bosonic action (\ref{Sbos1}) is invariant under the Abelian 
$U(1)$ gauge transformations (\ref{gauge}). We shall consider 
the supersymmetry and Lorentz invariance of the full
$N=2$,  $D=9$ ation in section 4  where the supersymmetry 
transformation rules for the fields will be derived.

Using the general expression  for dimensional reduction of the 
Einstein-Hilbert term (\ref{EH2}), one can rewrite the action 
(\ref{Sbos1}) in the $p$-brane metric \cite{duff} which
appears naturally in the $p$-brane $\sigma$-models and is 
related to the canonical gravitational matric in $D$ dimensions as 
$g_{\mu\nu}(p-brane)=e^{a/(p+1)}g_{\mu\nu}$, with 
\cite{LPSS}
\begin{equation}
a^2=\Delta-\fft{2(p+1)\tilde{d}}{D-2}, \label{a2}
\end{equation}
where $\tilde{d}=D-p-3$ and  $\Delta$ in maximal supergravity 
theories is equal to $4$. Then for the $p$-brane metric, the 
parameter $\gamma$  is defined by the equation: 
\begin{equation}
2\kappa (p+1)(7\gamma+1)(-2/\beta)^{1/2}=-(D-2)a.
\end{equation}
In the following, we use the canonical value of $\gamma$, which in 
$D=9$  is  $-1/7$.

\section{Fermionic sector}

We shall now compactify the fermionic part of the elven-
dimensional supergravity Lagrangian which reads \cite{CJS}
\begin{eqnarray}
{\cal L}_F&=&{\cal L}_F^{(1)}+{\cal L}_F^{(2)}+quartic 
        \  fermions, \\
{\cal L}_F^{(1)}&=&\fft{\h e}{2}{\h {\bar\psi_{\h r}}}
      {\h \Gamma}^{\h r\h s\h t}
                 {\h D}_{\h s}(\h \omega){\h \psi}_{\h t}\ , 
      \label{LF11}\\
{\cal L}_F^{(2)}&=&\fft{\kappa\h e}{98}
        \left({\h {\bar\psi_{\h r}}}
{\h \Gamma}^{\h r\h s\h t\h u\h v\h w}{\h \psi}_{\h s}            
+12{\h {\bar\psi^{\h t}}}{\h \Gamma}^{\h u\h v}{\h \psi}^{\h w}
\right)\h F_{\h t\h u\h v\h w}\ , 
\label{LF22}
\end{eqnarray}
where the covariant derivative is given by
\begin{equation}
{\h D}_{\h s} {\h \psi}_{\h t}=\del_{\h s} {\h \psi}_{\h t}
+\ft{1}{4}{\h \omega}_{\h s\h u\h v}
			{\h \Gamma}^{\h u\h v} {\h \psi}_{\h t}
+{\h \omega}_{\h s\h t}~^{\h w} {\h \psi}_{\h w}\ . 
     \label{hDPs} 
\end{equation}

The covariant derivative (\ref{hDPs}) commutes with the 
$\Gamma$-matrices: $[{\h D}_{\h s},{\h \Gamma}_{\h r}]=0$, 
which obey the algebra
\begin{equation}
[{\h \Gamma}_{\h r},{\h \Gamma}_{\h s}]=2\eta_{{\h r}{\h s}},
\end{equation}
where $\eta_{{\h r}{\h s}}=diag(-++\cdots+)$. 

The unhatted fermions and $\Gamma$-matrices are defined as
 follows
\begin{eqnarray}
\psi_r= {\h \psi}_r, \ \ \  \psi_i= {\h \psi}_i\ , \nonumber\\
\Gamma_r=\h \Gamma_r,\ \ \ \Gamma_i=\h \Gamma_i\ . 
\label{hPs}
\end{eqnarray}
The fermions and the $\Gamma$-matrices with the world indices
are defined by $\h \Gamma_\mu={\h e}^{~\h r}_\mu\h 
\Gamma_{\h r}$ and
$ {\h \psi}_\mu={\h e}^{~\h r}_\mu {\h \psi}_{\h r}$, or using 
(\ref{veil}) and (\ref{hPs}),
\begin{eqnarray} 
\h \Gamma_\mu=e^{-\frac{2}{3\!\sqrt{7}}\kappa\phi}
\Gamma_\mu+
         2\kappa e^{\frac{\!\sqrt{7}}{3}\kappa\phi}
{\cal A}_\mu^i\Gamma_i, \\
 {\h \psi}_\mu=e^{-\frac{2}{3\!\sqrt{7}}\kappa\phi}\psi_\mu+
2\kappa e^{\frac{\!\sqrt{7}}{3}\kappa\phi}{\cal A}_\mu^i\psi_i.
\end{eqnarray}
In order to bring the reduced Lagrangian to the canonical form, we 
have to redefine the fermionic fields
\begin{eqnarray}
\psi_r & \longrightarrow & \left(\psi_r-\frac{1}{7}\Gamma_r
\Gamma^i\chi_i\right)
	e^{\frac{1}{3\!\sqrt{7}}\kappa\phi}\ , \nonumber\\
\psi_i & \longrightarrow & \chi_i e^{\frac{1}{3\!\sqrt{7}}
\kappa\phi}\ . \label{PsPs}
\end{eqnarray}
In the kinetic term for the fermions, for example, the exponential 
in (\ref{PsPs}) cancels against the corresponding factors coming 
from the determinant $\h e$ and the covariant derivative
${\h D}_{\h s}$, and the shift
of the fermionic field ensures that the Lagrangian is diagonalised.
In the arbitrary dimension $D$, one has to make the following 
redefinitions 
\begin{eqnarray}
\psi_r & \longrightarrow & \left(\psi_r-\frac{1}{D-2}
\Gamma_r\Gamma^i\chi_i\right)
	\delta^{-\frac{1}{2}(\gamma(D-1)+1)}, \\ 
\psi_i & \longrightarrow & \chi_i \delta^{-\frac{1}{2}
(\gamma(D-1)+1)}.
\end{eqnarray}

Substituting (\ref{hDPs}), (\ref{hPs}), (\ref{PsPs}) into 
(\ref{LF11}) and (\ref{LF22}), and using the identity
\begin{equation}
\Gamma_{[s}\Gamma^{r_1\cdots r_n}\Gamma_{t]}=
{{\Gamma_s}^{r_1\cdots r_n}}_t
+n(n-1)\delta_s^{[r_1}\Gamma^{r_2\cdots r_{n-1}}
\delta_t^{r_n]},
\end{equation}
we derive the fermionic part of the $N=2$, $D=9$ supergravity 
Lagrangian
\begin{eqnarray}
{\cal L}_F^{(1)} &=& \fft{e}{2}\bar\psi_\mu
\Gamma^{\mu\nu\rho}{\cal D}_\nu\psi_\rho
+\fft{e}{2}\bar\chi_i\Gamma^\mu\left(\fft{1}{7}
\Gamma^i\Gamma^j+\delta^{ij}\right)
{\cal D}_\mu\chi_j \cr
&+& \fft{\kappa e}{4}e^{\frac{3}{\!\sqrt{7}}\kappa\phi}
   {\cal F}^j_{\mu\nu}
          \Big[-\fft{1}{2}\bar\psi^\rho\Gamma_{[\rho}
\Gamma^{\mu\nu}\Gamma_{\sigma]}
           \Gamma^j\psi^\sigma
          +\bar\psi^\rho\Gamma^{\mu\nu}\Gamma_\rho
          \left(\fft{1}{7}\Gamma^j\Gamma^k+\delta^{jk}\right)
          \chi_k \cr
&&\ \ \ \ \ \ \ \ \ \ \ \ \ \ \ +\fft{1}{49}\bar\chi_i
\Gamma^{\mu\nu}
	\left(-23\Gamma^k\delta^{ij}+59\Gamma^j\delta^{ik}
\right)\chi_k \Big]\ , \label{LF1}
\end{eqnarray}
\begin{eqnarray}
{\cal L}_F^{(2)} &=& \fft{\kappa e}{96}e^{\frac{2}
{\!\sqrt{7}}\kappa\phi}F'_{\mu\nu\rho\sigma}
          \Big[\bar\psi^\lambda\Gamma_{[\lambda}
\Gamma^{\mu\nu\rho\sigma}\Gamma_{\tau]}\psi^\tau
+\fft{6}{7}\bar\psi_\lambda\Gamma^{\mu\nu\rho\sigma}
\Gamma^\lambda\Gamma^i\chi_i \cr
&&\ \ \ \ \ \ \ \ \ \ \ \ \ \ \ \ \ \ +\bar\chi_i
\Gamma^{\mu\nu\rho\sigma}\left(\fft{11}{49}\Gamma^i
    \Gamma^j-\delta^{ij}\right)\chi_j\Big]\cr
&+& \fft{\kappa e}{24}e^{-\frac{1}{\!\sqrt{7}}\kappa\phi}
F'_{\mu\nu\rho i}
       \Big[-\bar\psi^\lambda\Gamma_{[\lambda}
\Gamma^{\mu\nu\rho}\Gamma_{\sigma]}
      \Gamma^i\psi^\sigma+2\bar\psi_\lambda
\Gamma^{\mu\nu\rho}\Gamma^\lambda\left(\delta^{ij}-
                  \fft{2}{7}\Gamma^i\Gamma^j\right)\chi_j\cr
&&\ \ \ \ \ \ \ \ \ \ \ \ \ \ \ \ \ \ -\fft{6}{7}\bar\chi_k\
Gamma^{\mu\nu\rho}\Gamma^k\left(\delta^{ij}-
       \fft{8}{7}\Gamma^i\Gamma^j\right)\chi_j\Big]\cr
&+&\fft{\kappa e}{16}e^{-\frac{4}{\!\sqrt{7}}\kappa\phi}
F_{\mu\nu ij}
        \Big[\bar\psi^\lambda\Gamma_{[\lambda}\Gamma^{\mu\nu}
         \Gamma_{\sigma]}\Gamma^{ij}\psi^\sigma
         -\fft{24}{7}\bar\psi_\lambda\Gamma^{\mu\nu}
\Gamma^{\lambda}\Gamma^i\chi^j\cr
&&\ \ \ \ \ \ \ \ \ \ \ \ \ \ \ \ \ \ +2\bar\chi_k\Gamma^{\mu\nu}
\left(\fft{16}{49}\Gamma^k\Gamma^i
+\delta^{ki}\right)\chi_j\Big] \ , \label{LF2}   
\end{eqnarray}
where  the covariant derivetives  
${\cal D}_\mu\psi_\nu=e_\mu^{~s}e_\nu^{~t}{\cal D}_s\psi_t$ 
and ${\cal D}_\mu\chi_i=e_\mu^{~s}{\cal D}_s\chi_i$
are defined as
\begin{eqnarray}
{\cal D}_s\psi_t=\partial_s\psi_t+\ft{1}{4}\omega_{suv}
\Gamma^{uv}\psi_t             
                             +\omega_{st}~^u\psi_u+\ft{1}{4}Q_{sij}
\Gamma^{ij}\psi_t,\\
{\cal D}_s\chi_j=\partial_s\chi_j+\ft{1}{4}\omega_{suv}
\Gamma^{uv}\chi_j+
                             \ft{1}{4}Q_{sik}\Gamma^{ik}\chi_j
+Q_{sj}~^k\chi_k.          
\end{eqnarray}
In deriving (\ref{LF1}) and (\ref{LF2}), we have used the following 
flipping property of Majorana spinors in nine 
dimensions
\begin{equation}
\bar\psi \Gamma^{r_1\cdots r_n}\eta=(-1)^n\bar\eta 
\Gamma^{r_n\cdots r_1}\psi.
\end{equation}

\section {Supersymmety transformations}

In this section, we obtain the supersymmetry transfomation laws 
in nine dimensions. In order to preserve the triangular form of the 
veilbein $\h e_\alpha^{~r}=0$,  one has to consider combined 
Lorentz and supersymmetry transformation laws. As we shall see, 
the requirement of the off-diagonal part of the veilbein be 
zero, imposes an additional constrain on the Lorentz group 
parameters. This, in its turn,  affects the supersymetry 
transformation laws of the fields.

Combining the supersymmetry and the Lorentz tranformation laws 
in eleven dimensions, we have \cite{CJS}
\begin{eqnarray}
 \delta\h e_{\h \mu\h r}&=&-\bar\eta \Gamma_{\h r}
{\h \psi}_{\h \mu}
                                 +\Lambda_{\h r\h s}\h e_{\h \mu}^{~\h s}\ , 
\label{de}\\
\delta \h A_{{\h \mu}{\h \nu} {\h \rho}}&=&-\ft{3}{2}\bar\eta 
\h \Gamma_{[{\h \mu}{\h \nu}}
         \h \psi_{{\h \rho}]}\ , \label{dA}\\
\delta\h \psi_{\h \mu}&=&\h D_{\h \mu}\eta -\ft{1}{144}
\left(\h \Gamma^{{\h \nu}{\h \rho}
                    {\h \sigma}{\h \gamma}}_{~~~~~{\h \mu}}+
                  8\h \Gamma^{{\h \nu}{\h \rho}{\h \sigma}}
\delta_{\h \mu}^{\h \gamma}\right)
                  \h F_{{\h \nu}{\h \rho}{\h \sigma}{\h \gamma}}
\eta+
                  \ft{1}{4}\Lambda_{{\h r}{\h s}}\Gamma^{{\h r}
{\h s}}\ , \label{dPs}
\end{eqnarray}
where $\eta$ is a supersymmetry parameter, $\h D_{\h \mu}\eta$
is a torsion free covariant derivative, and 
$\Lambda_{{\h r}{\h s}}=-\Lambda_{{\h s}{\h r}}$ is the Lorentz
group parameter.

Redefining the fermionic fields according to (\ref{PsPs}) and
taking different projections of (\ref{de})-(\ref{dPs}), we obtain 
the corresponding transformations in nine dimensions. 
The off-diagonal ($r\alpha$) projection of (\ref{de}) fixes the 
($ri$) component of the Lorentz parameter in terms of the 
supersymmetry parameter as follows
\begin{equation}
\Lambda_{ri}=\bar\ve\Gamma_r\chi_i\ , \label{Lamb}
\end{equation}
where $\ve=\eta e^{\frac{1}{3\!\sqrt{7}}\kappa \phi}$.
Eq.(\ref{Lamb}) is the necessary condition for  the triangular 
gauge $\h e_\alpha^{~r}=0$ to be preserved. Naively, one might 
expect that the $\xi^\mu$-reparametrization invariance  alone 
would be sufficient to maintain the triangular gauge of the vielbein. 
However, the explicit calculation shows that the reparametrization
transformation drops out of the ($r\alpha$) projection of 
(\ref{de}) altogether and, therefore, cannot modify constrain 
(\ref{Lamb}) on the Lorentz parameter. Substituting (\ref{Lamb}) 
into the ($r\mu$) projection of (\ref{de}), we obtain the veilbein 
transformation law
\begin{equation}
e_s^{~\mu}\delta e_{\mu r}=\ft{2}{3\!\sqrt{7}}\kappa
\delta_{sr}(\delta\phi) 
-\bar\ve\Gamma_{(r}\psi_{s)}
+\ft{1}{7}\bar\ve\Gamma^j\chi_j\delta_{rs}+\Lambda'_{rs}\ , 
\label{ede}
\end{equation}
where symmetrization is performed with the unit strength  and  
$\Lambda'_{sr}$ is the 
redefined  local $SO(1,8)$ Lorentz transformation parameter 
\begin{equation}
\Lambda'_{sr}=\Lambda_{sr}-\bar\ve\Gamma_{[s}\psi_{r]}+
                        \ft{1}{7}\bar\ve\Gamma_{rs}\Gamma^j\chi_j.
\end{equation}

Remaining two projections of (\ref{de}) give the transformation 
rules for the vector field 
${\cal A}_\mu^i$ and the internal veilbein $L_\alpha^{~i}$:
\begin{eqnarray}
\delta {\cal A}_\mu^i&=&-\ft{\!\sqrt{7}}{3}\kappa 
{\cal A}_\mu^i\delta \phi 
                   -\ft{1}{2\kappa}\bar\ve e^{-\frac{3}{\!\sqrt{7}}
\kappa\phi}
            \left(\Gamma^i\psi_\mu +\Gamma_\mu(\delta_{ij}
+\ft17\Gamma^i\Gamma^j)\chi_j\right)\cr
 &&-\bar\ve\Gamma_{(i}\chi_{j)}{\cal A}_\mu^j
                       + \Lambda'^i_{~j}{\cal A}_\mu^j,\\
L_j^{~\alpha}\delta L_{\alpha i}&=&-\ft{\!\sqrt{7}}{3}\kappa
\delta_{ij}(\delta\phi) -
\bar\ve\Gamma_{(i}\chi_{j)}+\Lambda'_{ij}\ , 
\label{LdL} 
\end{eqnarray}
where the redefined $SO(2)$ Lorentz parameter is
\begin{equation}
\Lambda'_{ij}=\Lambda_{ij}-\bar\ve\Gamma_{[i}\chi_{j]}.
\end{equation}
Tracing (\ref{LdL}) with $\delta_{ij}$, one finds the 
transformation law for $\delta\phi$, which upon 
substitution into (\ref{ede}), puts the veilbein transformation 
law into canonical form. Suppressing
the $\Lambda'_{rs}$ and $\Lambda'_{ij}$ transformations, 
we obtain the supersymmetry transformation 
rules for the bosonic fields:
\begin{eqnarray}
&&\cr
\delta e_\mu^{~r}&=&-\bar \ve\Gamma^r\psi_\mu,\\
&&\cr
\delta \phi&=&-\ft{3}{2\!\sqrt{7}\kappa}\bar\ve\Gamma^i
\chi_i,\\
&&\cr
L_j^{~\alpha}\delta L_{\alpha i}&=&-\bar\ve
\left(\Gamma_{(i}\chi_{j)}-
\ft{1}{2}\delta_{ij}\Gamma_k\chi_k\right),\\
&&\cr
\delta {\cal A}_\mu^i&=&{\cal A}_\mu^j\left(L_j^{~\alpha}
\delta L_{\alpha i}\right)
-\ft{1}{2\kappa}\bar\ve e^{-\frac{3}{\!\sqrt{7}}k\phi}
\left(\Gamma^i\psi_\mu +\Gamma_\mu(\delta_{ij}
+\ft17\Gamma^i\Gamma^j)\chi_j\right),\\
&&\cr
\delta A_{\mu\alpha\beta}&=&-\ft{1}{2}\bar\ve 
e^{\frac{4}{\!\sqrt{7}}\kappa\phi}
             \Gamma_{\alpha\beta}\psi_\mu
-\bar\ve  e^{\frac{4}{\!\sqrt{7}}\kappa\phi}\left(\ft{8}{7}
\Gamma_\mu\Gamma_{[\alpha}
 +3\kappa e^{\frac{3}{\!\sqrt{7}}\kappa\phi}
{\cal A}_\mu^\gamma
\Gamma_{[\gamma\alpha}\right)\chi_{\beta]},\\
&&\cr
\delta A_{\mu\nu\alpha}&=&-\ft{1}{2}\bar\ve  
e^{\frac{1}{\!\sqrt{7}}\kappa\phi}
\left(\Gamma_\alpha\Gamma_{[\mu}-2\kappa 
e^{\frac{3}{\!\sqrt{7}}\kappa\phi}  
\Gamma_{\alpha\gamma}{\cal A}_{[\mu}^\gamma\right)
\psi_{\nu]}\cr
  &&-\ft{1}{2}\bar\ve e^{\frac{1}{\!\sqrt{7}}\kappa\phi} 
\Gamma_{\mu\nu}\left(
                   \delta_\alpha^\beta-\ft{2}{7}\Gamma_\alpha
\Gamma^\beta\right)\chi_\beta\cr
 &&-4\kappa\bar\ve   e^{\frac{4}{\!\sqrt{7}}\kappa\phi}
\left(\ft{6}{7}{\cal A}_{[\mu}^\gamma
\Gamma_{\nu]}\Gamma_{[\alpha}\delta_{\gamma]}^\beta
+\kappa e^{\frac{3}{\!\sqrt{7}}\kappa\phi} 
{\cal A}_{[\mu}^\gamma{\cal A}_{\nu]}^\delta
\Gamma_{\gamma [\delta}\delta_{\alpha]}^\beta\right)
\chi_\beta,\\
&&\cr
\delta A_{\mu\nu\rho}&=&-\ft{3}{2}\bar\ve e^{-\frac{2}
{\!\sqrt{7}}\kappa\phi}\Gamma_{[\mu\nu}\psi_{\rho]}
        +\ft{3}{14}\bar\ve e^{-\frac{2}{\!\sqrt{7}}\kappa\phi}
\Gamma_{\mu\nu\rho}\Gamma^\a
               \chi_\a\cr
&&+6\bar\ve e^{\frac{1}{\!\sqrt{7}}\kappa\phi}\left(
\Gamma_\alpha\Gamma_{[\mu}{\cal A}_\nu
-\kappa e^{\frac{3}{\!\sqrt{7}}\kappa\phi}
\Gamma_{\alpha\beta}{\cal A}_{[\mu}^\alpha
 {\cal A}_\nu^\beta\right)\psi_{\rho]}\cr
&&-3\kappa\bar\ve e^{\frac{1}{\!\sqrt{7}}\kappa\phi}
\left(\Gamma_{[\mu\nu}{\cal A}_{\rho]}^\alpha
          (\delta_\alpha^\gamma-\ft{2}{7}\Gamma_\alpha
\Gamma^\gamma)
+\ft{24}{7}\kappa e^{\frac{3}{\!\sqrt{7}}\kappa\phi}
{\cal A}_{[\mu}^\alpha{\cal A}_\nu^\beta
\Gamma_{\rho]}\Gamma_\alpha\delta_\beta^\gamma\right)
\chi_\gamma\cr
&&
\end{eqnarray}

Using (\ref{hDPs}), (\ref{hPs}) and (\ref{PsPs}), and 
performing the straightforward reduction of (\ref{dPs}),
one obtaines the transformation laws for the fermionic fields:
\begin{eqnarray}
\delta\psi_\mu&=&{\cal D}_\mu\ve-\ft{1}{28}\kappa 
e^{\frac{3}{\!\sqrt{7}}\kappa\phi}
 {\cal F}^i_{\nu\rho}\left(
                        \Gamma_\mu~^{\nu\rho}+12\Gamma^\nu
\delta_\mu^\rho\right)\Gamma^i\ve\cr
 &&- \ft{1}{144}e^{\frac{2}{\!\sqrt{7}}\kappa\phi}
F'_{\nu\rho\sigma\lambda}\left(\ft{9}{7}
\Gamma^{\nu\rho\sigma\lambda}_{~~~~\mu}+
\ft{48}{7}\Gamma^{\nu\rho\sigma}
         \delta_\mu^\lambda\right)\ve\cr
 &&+ \ft{1}{42}e^{-\frac{1}{\!\sqrt{7}}\kappa\phi}
F'_{\nu\rho\sigma j}\left(
     			\Gamma^{\nu\rho\sigma}_{~~~\mu}+\ft{15}{2}
\Gamma^{\nu\rho}\delta_\mu^\sigma\right)
          \Gamma^j\ve\cr
&&- \ft{1}{7\times24}e^{-\frac{4}{\!\sqrt{7}}\kappa\phi} 
F_{\nu\rho ij}
         \left(3\Gamma^{\nu\rho}_{~~\mu}
      +36\Gamma^\nu\delta^\rho_\mu
      \right)\Gamma^{ij}\ve \ + \ cubics\ , \cr
&&\cr
\delta\chi_i&=&-\ft{1}{4}\kappa e^{\frac{3}{\!\sqrt{7}}\kappa
\phi}{\cal F}^i_{\mu\nu}\Gamma^{\mu\nu}\ve
+(P_{\mu ij}+\ft{\sqrt{7}}{3}\kappa\delta_{ij} 
\partial_\mu\phi)\Gamma^\mu\Gamma^j\ve \cr
 &&-\ft{1}{144}e^{\frac{2}{\!\sqrt{7}}\kappa\phi}
F'_{\mu\nu\rho\sigma}\Gamma^{\mu\nu\rho\sigma}
\Gamma_i\ve+\ft{1}{36}e^{-\frac{1}{\!\sqrt{7}}\kappa\phi}
F'_{\mu\nu\rho j}\Gamma^{\mu\nu\rho}(\Gamma_{ij}
    -2\delta_{ij})
                  \ve \cr
&&-\ft{1}{6}e^{-\frac{4}{\!\sqrt{7}}\kappa\phi}
F_{\mu\nu ji}\Gamma^{\mu\nu}\Gamma^j\ve
         \ +\ cubics \ ,
\end{eqnarray}
where
\begin{equation}
{\cal D}_\mu\ve=\partial_\mu\ve+\ft{1}{4}\omega_{\mu st}
\Gamma^{st}\ve+ \ft{1}{4}Q_{\mu ij}\Gamma^{ij}\ve
\end{equation}

\section{Stainless  $p$-brane solutions}

Most supergravity theories in $D<11$  dimensions can be obtained 
by dimensionally reducing $D=11$ supergravity theory. 
Therefore,  solutions of the dimensionally reduced theory  are also
solutions of the higher-dimensional theory. However, in higher 
dimension these solutions may or may not exhibit isotropicity.  The 
solutions which cannot be isotropically lifted to the higher 
dimension and, therefore, cannot be viewed as $p$-brane solutions 
in the higher dimensions, are  called stainless solutions 
\cite{LPSS}.  In other words, stainless $p$-branes  are genuinely 
new solutions of the supergravity theory in the given 
dimension and should not be treated on the same footing as solutions 
which are descendants of the higher dimensional $p$-branes.

We first briefly review some of the main results on $p$-brane 
solutions \cite{duff,LPmax,LPSS} and then apply general results 
to constructing and classifying stainless $p$-branes in $N=2$, 
$D=9$
theory.
The $p$-brane solutions in general involve the metric tensor 
$g_{\sst {MN}}$, a dilaton $\phi$ and an $n$-index 
antisymmetric tensor $F_{\sst {M_1M_2\cdots M_n}}$. 
The Lagrangian for these fields takes the form
\begin{equation}
e^{-1}{\cal L}=R-\ft12 (\del\phi)^2-\ft{1}{2n!}e^{-a\phi}
F_n^2\ , \label{L}
\end{equation}
where $a$ is a constant given by (\ref{a2}) \cite{duff,LPSS}.  
In $D=11$, the absence of a dilaton implies that $\Delta=4$.  
The value of $\Delta$ is preserved under the dimensional 
reduction procedure and, hence, all antisymmetric tensors in 
maximal supergravity theories have $\Delta=4$.
However, if an antisymmetric tensor used in a particular 
$p$-brane solution is formed from a linear combination of the 
original field strengths, then it will have $\Delta<4$. An example 
of this is a solitonic 5-brane in $N=1$, $D=10$ supergravity 
considered below.

We shall be looking for isotropic $p$-brane solutions for which 
the metric ansatz is given by \cite{duff,LPSS}
\begin{equation}
ds^2=e^{2A}dx^\mu dx^\nu\eta_{\mu\nu}+e^{2B}dy^m dy^m,
\end{equation}
where $x^\mu $($\mu=0,\cdots ,d-1$) are the coordinates of the 
$(d-1)$-brane world volume, and $y^m$ are the coordinates of 
the $(9-d)$-dimensional transverse space. The functions A and B, 
also the dilaton $\phi$, depend only on $r=\sqrt{y^my^m}$. This 
ansatz for the metric preserves an $SO(1, d-1)\times SO(9-d)$ 
subgroup of the original $SO(1,8)$ Lorentz group.

 For the elementary $p$-brane solutions, the ansatz for the field 
strenght  is given by \cite{duff,LPSS}
\begin{equation}
F_{m\mu_1\cdots\mu_{n-1}}=\ve_{\mu_1\cdots\mu_{n-1}}
\del_me^C, \label{elem}
\end{equation}
where $\ve_{\mu_1\cdots\mu_{n-1}}\equiv g_{\mu_1\nu_1}
\cdots\ve^{\nu_1\cdots}$ with $\ve^{012\cdots}=1$, 
and C is a function of $r$ only. The dimension of the brane 
world volume is $d=n-1$.

For the solitonic $p$-brane solution, the ansatz for the 
antisymmetric tensor is given by \cite{duff,LPSS}
\begin{equation}
F_{m_1\cdots m_n}=\lambda\ve_{m_1\cdots m_np}
\frac{y^p}{r^{n+1}}, \label{solit}
\end{equation}
where $\lambda$ is a constant and the dimension of the world 
volume is $d=9-n-1$.

The solutions to the equations of motion obtained from the 
Lagrangian (\ref{L}) are given by
\begin{eqnarray}
A&=&-\fft{2\tilde{d}}{\Delta (D-2)} ln\left(1+\fft{k}
{r^{\tilde{d}}}\right), \cr
B&=&-\fft{d}{\tilde{d}}A, \quad \phi=\fft{7\e a}{\tilde{d}}A \ , 
\label{sol}
\end{eqnarray}
where 
\begin{equation}
k=\fft{\e \lambda}{2\tilde{d}}\sqrt{\Delta}\ , \nonumber
\end{equation}
$\tilde{d}=D-d-2$ and $\e=1$ ($\e=-1$) for the elementary 
(solitonic) ansatz. In the solitonic case, the equation of motion for 
the field strength is automatically satisfied, whilst in the 
elementary case the function $C$ is given by
\begin{equation}
e^C=\fft{2}{\sqrt{\Delta}} {\left(1+\fft{k}{r^{\tilde{d}}} 
\right)}^{-1}\ . \label{eC}
\end{equation}

The solutions (\ref{sol})-(\ref{eC}) are valid for an $n$-index 
fild strength with $n>1$. When $n=1$,  i.e.  $\tilde{d}=0$ , there 
only exists a solitonic solution described by (\ref{sol}) with 
$kr^{-\tilde{d}}\ra klogr$ and $\tilde{d}\ra 0$.

Stainlessness  of a $p$-brane solution  crucially depends on a 
degree of the antisymmtric tensor involved in a  solution, and the 
value of constant $a$ occurring in the exponential prefactor.
There are two different situations when a stainless $p$-brane 
solution may arise in a given dimension. In the first scenario, 
no $(D+1)$-dimensional theory contains the necessary field 
strength for a brane solution. In particular, if in $D$ dimensions 
the solution is elementary, the $(D+1)$-dimensional
theory must have a field strength of degree one higher than that 
in $D$ dimensional theory. If it is a solitonic solution, the 
$(D+1)$-dimensional theory must contain a field strength of 
the same degree as in $D$ dimensional theory. In the second case, 
the required field strength exists in the $(D+1)$-dimensional 
theory, but a $p$-brane is stainless only if the constant $\h a$ 
of a corresponding antisymmtric tensor in $(D+1)$ dimensions 
is not related to the constant $a$ of the $D$-dimensional theory
as \cite{LPSS}
\begin{equation}
{\h a}^2=a^2-\frac{2\tilde{d}^2}{(D-1)(D-2)} \label{arel}
\end{equation}

It should be noted that in determining whether or not a particular 
$p$-brane solution is stainless, we restrict our attention only to 
the supergravity theories which can be obtained from $D=11$
supergravity. This, for example, will lead us to conclude that a 
solitonic 6-brane and an elementary membrane in the nine-dimensional 
theory are stainless. However, if we include type-IIB 
theory into consideration, we will see that these $p$-branes are no 
longer stainless and can be isotropically oxidized to the solitonic 
7-brane and the self-dual 3-brane of type IIB supergravity.

In order to consider solutions to the obtained $N=2$, $D=9$ 
supergravity theory, we need to parametrize tensor 
$P_{\mu ij}$ in the Lagrangian (\ref{Sbos1}).  Since after 
separating out the determinant, the internal veilbein has only two 
degrees of freedom left, we introduce two scalar fields,
$\vp$ and $E$, and parametrize the veilbeins as follows:
\begin{equation}
 L_{\a i}=
\pmatrix{e^{\kappa\vp}&\kappa Ee^{-\kappa\vp}\cr 
0                                      &e^{-\kappa\vp}\cr},\qquad
 L^{\a}_i=
\pmatrix{e^{-\kappa\vp}&-\kappa Ee^{-\kappa\vp}\cr 
                     0                   & e^{\kappa\vp}\cr}\ .
\end{equation}
The metric $g_{\a\b}$ is given by
\begin{equation}
g_{\a\b}=\pmatrix{e^{2\kappa\vp}+\ft14 E^2 e^{-2\kappa\vp}
&        \ft12 e^{-2\kappa\vp}\cr
\ft12 e^{-2\kappa\vp} &e^{-2\kappa \vp}\cr}. \label{intlg}
\end{equation}
The internal metric (\ref{intlg}) is not diagonal, therefore, the 
terms in the Lagrangian containing $g_{\alpha\beta}$ will not 
be diagonal as well.  To diagonalize the Lagrangian, the following 
redefinitions
have to be made:
\begin{eqnarray}
{\cal F}_{\sst {MN}}^{2}+ E{\cal F}_{\sst {MN}}^{1}\ra  
{\cal F}_{\sst {MN}}^{(2)}, && \qquad
{\cal F}_{\sst {MN}}^{1}\ra  {\cal F}_{\sst {MN}}^{(1)}, \cr
&&\cr
{F'}_{\sst {MNP}}^{2}+ E {F'}_{\sst {MNP}}^{1}\ra   
{F'}_{\sst {MNP}}^{(2)}, && \qquad
{F'}_{\sst {MNP}}^{1}\ra   {F'}_{\sst {MNP}}^{(1)}\ . 
\end{eqnarray}
Here and throughout this subsection   $M,N,P=0,\cdots ,8$ denote 
the 
curved nine-dimensional world volume indices, whilst 
$R,S,T=0,\cdots ,8$ denote the flat indices.

Then the Lagrangian (\ref{Sbos1}) can be written as
\begin{eqnarray}		
e^{-1}{\cal L}&=&R-\ft12 (\del\phi)^2-\ft12 (\del\vp)^2 
-\ft14 e^{\ft{3}{\sqrt{7}}\phi+\vp}({\cal F}_2^{(1)})^2
-\ft14 e^{\ft{3}{\sqrt{7}}\phi-\vp}({\cal F}_2^{(2)})^2\cr
 &&-\ft12 e^{2\vp}(H_1)^2+\ft18 e^{-\ft{4}{\sqrt{7}}\phi}
(F_2)^2
-\ft1{12}e^{-\ft1{\sqrt{7}}\phi+\vp}{(F'^{(1)}_3)}^2\cr
&&+\ft1{12}e^{-\ft1{\sqrt{7}}\phi-\vp}{(F'^{(2)}_3)}^2
-\ft1{48}e^{\ft2{\sqrt{7}}\phi}(F'_4)^2+{\cal L}_{FFA}\ , 
\label{Sbos2}
\qquad
\end{eqnarray}
where 
$$F_n^2\equiv F_{\sst {M_1M_2\cdots M_n}}
F^{\sst {M_1M_2\cdots M_n}}$$ 
denotes the square of an n-index
field strength, $H_{\sst M}=\del_{\sst M} E$ and the parameter 
$\kappa$ is set to $1/2$. 

If  in (\ref{Sbos2}) one retains only one field strength and a 
corresponding dilaton, which for ${\cal F}_2^{(i)}$ and 
$F'^{(i)}$ is a linear combination of  $\phi$ and $\vp$,
one arrives at the Lagrangian of the form (\ref{L}). Then general 
results described above can be applied to constructing single 
$p$-brane solutions in the given supergravity theory.

It should be noted that for the purposes of finding a purely 
elementary or a purely solitonic $p$-brane solution,  the 
$FFA$-term in the Lagrangian and the Chern-Simons 
modifications of the field strengths can be disregarded due to the 
fact that the constraints implied by these terms are 
automatically satisfied in $D=9$. However, in general, for 
certain $p$-brane solutions, the ${\cal L}_{FFA}$ term and 
the Chern-Simons modifications to the field strengths give 
rise to nontrivial equations in some dimensions \cite{LPmax}. 
Good examples illustrating this point are dyonic $p$-branes in 
$D=4$ and $D=6$ dimensions.  

Since our principal interest lies in the stainless $p$-brane 
solutions, we have first to determine which of the $p$-branes of 
$N=2, D=9$ supergravity are stainless. For this, one recalls
that  the $N=2, D=10$ supergravity contains a 2-index field 
strength, a 3-index field strength and a 4-index field stregth with 
the ${\h a}^2$ values $1,\ft14,\ft94$ respectively. Using
the criteria for stainlessness of a $p$-brane and the equation 
(\ref{arel}), we find that the stainless solutions of $N=2, D=9$ 
supergravity theory are an elementary  particle,  an elementary 
membrane and a solitonic 6-brane.
Applying (\ref{sol}), we obtain the metrics for these solutions 
\begin{eqnarray}
{\rm  particle}:&& ds^2 =  \Big(1+\fft{k}{r^{6}}\Big)^{-6/7} 
dt^2+\Big(1+\fft{k}{r^{6}}\Big)^{2/7} dy^mdy^m \ ,
\label{part}\\
&&\cr 
{\rm   membrane}:&& ds^2 =\Big (1+\fft{k}{r^{4}}\Big)^{-4/7} 
dx^\mu dx^\nu\eta_{\mu\nu} + \Big(1+\fft{k}{r^{4}}
\Big)^{3/7} dy^mdy^m, \label{memb}\\
&&\cr 
{\rm  6-brane}:&& ds^2 = dx^\mu dx^\nu
\eta_{\mu\nu} + \Big(1+klogr\Big)^{3/7} dy^mdy^m\ . 
\label{6br}
\end{eqnarray}
It is of interest to note that a solitonic 5-brane, which is stainless 
as a solution to $N=1, D=9$ supergravity theory, does not remain 
stainless in $N=2, D=9$ supergravity. This seeming paradox 
can be resolved if we consider details of the truncation of $N=2$  
to $N=1$ supergravity, which contains a dilaton, a 2-index field 
strength and a 3-index field strength.  One cannot consistently 
truncate out either two 2-index antisymmetric tensors or a scalar 
field.  Nonetheless, it is possible to make a consistent truncation if 
we first rotate the scalar fields:
\begin{equation}
\varphi = \sqrt{\ft78} \phi_1 -\sqrt{\ft18} \phi_2\ ,\qquad
\phi = \sqrt{\ft18} \phi_1 + \sqrt{\ft78} \phi_2
\end{equation}
and then set 
$\phi_2=F_4=H_1=F_3^{(1)}={\cal F}_2^{(1)}=0$ which 
now is consistent with the equations of motion. Defining 
$\tilde{F_2}\equiv\sqrt{2}F_2=\sqrt{2} {\cal F}_2^{(2)}$, 
we get the Lagrangian for the bosonic sector of $N=1, D=9$ 
supergravity \cite{LPSS,GNS}:
\begin{equation}
{\cal L} = e R -\ft12 e(\del \phi_1)^2 -\ft1{12} e 
e^{-\sqrt{\fft87}\phi_1}{(F_3^{(2)})}^2 -\ft14 e 
e^{-\sqrt{\fft27}\phi_1} {\tilde{F}_2}^2\ .
\end{equation} 
We now see that to obtain a 5-brane solution described by  
$\tilde{F_2}$ in $N=1$ theory, one has to start with a multi-brane 
solution \cite{NZ} in $N=2$ supergravity, namely, a 
two-scalar 5-brane described by  $F_2$ and 
${{\cal F}_2}^{(2)}$. The metric for this solution, which 
preserves quarter of
the supersymmetry of $N=2$ supergravity, is given by
\begin{equation}
ds^2 =\Big (1+\fft{k_1}{r}\Big)^{-1/7}
\Big (1+\fft{k_2}{r}\Big)^{-1/7} dx^\mu dx^\nu
 \eta_{\mu\nu} + \Big (1+\fft{k_1}{r}\Big)^{6/7}
\Big (1+\fft{k_2}{r}\Big)^{6/7} dy^mdy^m\ .
\end{equation}
For $k_1=k_2$, the two-scalar 5-brane can also be viewed as a 
single solitonic 5-brane of $N=1$ supergravity preserving half of 
the supersymmetry of this theory. It should also be remarked 
that the $\Delta$ value for the $N=1$ 5-brane is 2, as opposed 
to $\Delta=4$ for the $N=2$ 5-brane, which is yet another 
indication of a multi-brane origin of this solution.

Thus we see that an interesting phenomenon of supersymmetry 
enhancement may occur when a single-brane solution of truncated 
theory can be viewed as a particular limit of a multi-brane  
solution of extended supergravity. Notice that not only does one 
obtain a solution preserving more supersymmetries, but one may 
also find a new stainless $p$-brane in the truncated 
supergravity.  

We now begin to examine in detail the stainless solutions 
(\ref{part})-(\ref{6br}) and varify that they  
preserve half of the supersymmetry.  To consider the 
supersymmetry of the elementary particle
solution (\ref{part}), we make a $1+8$ split of the gamma 
matrices:
\begin{equation}
\Gamma^0=\gamma_9,\quad \Gamma^m=\gamma^m,
\end{equation}
where $\gamma_9=\gamma_1\gamma_2\cdots\gamma_8$ and 
$\gamma_m$ are numerical
matrices with flat indices. The transformation rules for the 
fermions become
\begin{eqnarray}
\delta \chi &=& -\ft{\sqrt7}{6}\, e^{-B} \del_m \phi\, 
\Gamma_i\gamma_m \ve
- \ft13\, e^{-A - B+C - \fft{2}{\sqrt7}\phi}\, \del_m C \,
\gamma_m\gamma_9 \e_{ij}\Gamma^j\ve\ ,\nonumber\\
\delta \psi_0 &=& \ft12 e^{A-B}\, \del_m A\, \gamma_m
\gamma_9 \ve
-\ft{3}{14}\, e^{-B+C-\fft{2}{\sqrt7}\phi}\, \del_m C
\,\gamma_m\e_{ij}\Gamma^{ij} \ve\ ,\label{ferpart}\\
\delta \psi_m &=& \del_m \varepsilon +\ft12 \del_n B\, 
\gamma_{mn}\
\ve+ \ft{1}{28}\, e^{-A +C-\fft{2}{\sqrt7}\phi}\, \del_m C\,
\gamma_{mn}\gamma_9 \e_{ij}\Gamma^{ij}\ve\nonumber\\
&& +\ft{3}{14}\, e^{-A+C  -\fft{2}{\sqrt7}\phi} \,
\del_m C \,\gamma_9 \e_{ij}\Gamma^{ij}\ve\ .\nonumber
\end{eqnarray}
Substituting the solution (\ref{sol}) into the equations above, we 
find that the variations af all fermionic fields vanish provided that
\begin{equation}
\ve= e^{-\ft12 A}\, \ve_0\ ,\qquad
\gamma_9\,  \ve_0  =\ve_0\ ,\qquad
\e_{ij}\Gamma^{ij}\ve_0=\ve_0,
\end{equation}
where $\ve_0$ is a constant spinor. Thus the elementary particle 
solution preserves half of the supersymmetry.

To varify that the elementary membrane solution also preserves 
half of the supersymmetry, we make a $3+9$ split of the gamma 
matrices:
\begin{equation}
\Gamma^{\mu} = \gamma^\mu \otimes \gamma_7\ ,\qquad
\Gamma^{m} = \oneone \otimes \gamma^m \ ,
\end{equation}
where $\gamma_7 = \gamma_0\gamma_1\ldots\gamma_5$ in the 
transverse space and $\gamma_1\gamma_2\gamma_3=\oneone$ 
on the world volume.  The transformation rules for the fermions 
become
\begin{eqnarray}
\delta \chi_i &=& -\fft{\sqrt7}{6}\, e^{-B}\, \del_m \phi\, 
\gamma_7\otimes\gamma_m \Gamma_i\ve- \ft16\,
    e^{ -B-3A+C+\fft1{\sqrt7}\phi}\, \del_mC\, \oneone \otimes 
\gamma_m \Gamma_i\ve \ ,\nonumber\\
\delta \psi_\mu&=&  \del_m A\, e^{A-B}\, \gamma_\mu \otimes
\gamma_7\gamma_m \ve + \ft27\, e^{ -B-3A+C+\fft1{\sqrt7}
\phi}\, \gamma_\mu\otimes\gamma_7 \gamma_m \ve\ ,
\label{fermem}\\
\delta \psi_m &=& \del_m \ve + \ft12 \del_n B
            \oneone\otimes \gamma_{mn}\ve +
           \ft3{28}\,  e^{ -B-3A+C+\fft1{\sqrt7}\phi}\, \del_nC
           \oneone\otimes\gamma_{mn} \ve \nonumber\\
 &&+\ft27\, e^{ -B-3A+C+\fft1{\sqrt7}\phi}\, \del_mC \, \ve\ .
\nonumber
\end{eqnarray}

This solution preserves half of the supersymmetry provided that
\begin{equation}
\ve=\, e^{-A}\ve_0, \qquad
\gamma_7\otimes\oneone\, \ve_0=\ve_0\ .
\end{equation}

It can be shown that a solitonic 6-brane solution (\ref{6br}) also 
preserves half of the supersymmetry \cite{LPSS}.  This solution, 
as was discussed above, is stainless in $N=2$, $D=9$ 
supergravity, the reason being the absence of the required 
one-index field strength in type IIA  $D=10$ supergravity 
\cite{LPSS}. However, in $D=10$ there also exists the chiral 
type IIB supergravity theory which, unlike type IIA version, 
cannot be obtained by compactification of the eleven-dimensional 
$N=1$ supergravity, and which contains the necessary
one-index field strength.  Recalling that the coefficient $a$ in 
the exponential prefactor of this field stregth is 2, and that 
$\tilde{d}=0$, one concludes, based on (\ref{arel}), that the 
solitonic 6-brane of $N=2$, $D=9$ supergravity can be 
isotropically oxidized to the type IIB solitonic 7-brane.  A similar 
situation is encountered with the elementary membrane solution 
which turns out to be a descendant of the type IIB self-dual 
elementary 3-brane described by  a five-index
self-dual field strength with $a=0$. Thus, if  besides eleven-dimensional 
supergravity one is to include type IIB supergravity 
in the classification of $p$-brane solutions, one necessarily 
arrives at a conlcusion that the elementary particle is the only 
stainless solution in $N=2$, $D=9$ supergravity theory.

\section{Conclusions}

In this paper, we constructed and studied the extended $N=2$ 
supergravity theory in $D=9$. The full action was obtained by 
compactifying $D=11$ supergravity directly to $D=9$. The 
method employed was the ordinary Scherk-Schwarz dimensional 
reduction procedure which gives an advantage of 
constructing the lower dimensional theory in one step, as opposed 
to the standard Kaluza-Klein step by step dimensional reduction, 
and also enables one to consider compactification on a non-trivial 
group manifolds \cite{sez}. We explored the stainless $p$-brane 
solutions to the obtained $N=2$ supergravity in
$D=9$. Having derived the supersymmetry transfornation laws 
for the fields, we were in the position to examine the 
sypersymmetry of the found stainless $p$-branes. Discussing 
the relation of the $N=2$ solutions to the $N=1$ stainless 
solutions, it was observed that the stainless solutions of truncated 
theory may or may not remain stainless in the extended 
supergravity.  The notion of stainlessness was discussed  
in the case when, along with $D=11$ supergravity, type IIB 
supergravity was taken into consideration.

\section*{Acknowledgements}

We are grateful to E. Sezgin, C.N. Pope, H. Lu and M.J. Duff 
for useful discussions.


\begin{thebibliography}{20}
\frenchspacing
\bibitem{berg2} E. Bergshoeff, C. Hull and T. Ortin, 
{\it Duality in the type-II superstring effective action}, 
QMW-PH-95-2, hep-th/9504081.
\bibitem{das} A. Das and S. Roy, {\it On M-theory and the 
symmetries of type II string effective actions}, Nucl. Phys. 
{\bf B482} (1996) 119, hep-th/9605073.  
\bibitem{CJS} E. Cremmer, B. Julia and J. Scherk, 
{\it Supergravity theory in eleven dimensions}, 
              Phys. Lett. {\bf B76} (1978) 409.
\bibitem{rom} L. Romans,  {\it Massive $N=2a$ supergravity in 
ten dimensions}, Phys. Lett. {\bf 169B} (1986) 374.
\bibitem{berg1}E. Bergshoeff, M. De Roo, M.B. Green, 
G. Papadopoulos and \\ 
P.K. Townsend, {\it Duality of type II 
7-branes and 8-branes}, UG-15-95, hep-th/9601150.
\bibitem{SS}J. Scherk and J.H. Schwarz, {\it How to get masses 
from extra dimensions}, {\em Nucl. Phys.} {\bf B153} (1979) 61.
\bibitem{LPmass} P.M. Cowdall, H. Lu, C.N. Pope, K.S. Stelle 
and P.K. Townsend, {\it Domain walls in massive supergravities}, 
CTP-TAMU-26-96A, hep-th/9608173. 
\bibitem{duff} M.J. Duff, R. Khuri, J.X. Lu, Phys. Rept. 
{\it String solitons} {\bf 259} (1995) 213.
\bibitem{NZ} N. Khviengia, Z. Khviengia, H. Lu and C.N. Pope, 
{\it Intersecting M-branes and bound states}, Phys. Lett. 
{\bf B388} (1996) 21, hep-th/9605077.
\bibitem{LPmax} H. Lu and C.N. Pope, {\it p-brane solitons in 
maximal supergravities}, Nucl. Phys. {\bf B465} (1996) 127,  
hep-th/9512012.
\bibitem{LPSS}H. Lu, C.N. Pope, E. Sezgin and K.S. Stelle, 
{\it Stainless super p-branes}, Nucl. Phys. {\bf B456}, (1995) 
669, hep-th/9508042.
\bibitem{sez} A. Salam and E. Sezgin, {\it $d=8$ supergravity}, 
Nucl. Phys. {\bf B258} (1985) 284.
\bibitem{LPmulti}  H. Lu and C.N. Pope, {\it Multi-scalar p-brane 
solitons}, CTP-TAMU-52/95, hep-th/9512153.
\bibitem{GNS} S.J. Gates, H. Nishino and E. Sezgin, Class. Quantum 
Grav. {\bf 3} (1986) 21.
\end{thebibliography}
\end{document}